# Socio-economic and Technological Factors Influencing Financial Inclusion among Indigenous Peoples in Bauchi State, Nigeria

A., Hassan *

Department of Information Systems, School of Information Technology and Computing, American University of Nigeria, Yola, Nigeria, abduljalal.hasssan@aun.edu.ng

S. C. A. Utulu

Department of Information Systems, School of Information Technology and Computing, American University of Nigeria, Yola, Nigeria, samuel.utulu@aun.edu.ng

The need to understand the factors that come to bear in the financial inclusion on the indigenous peoples in Nigeria necessitated the study. The need is pressing because scholars have established that the financial inclusion is crucial to the socio-cultural and economic development of the indigenous peoples. Therefore, the study was informed by the research study question, what are the barriers to financial inclusion of the indigenous peoples in Nigeria? We adopted the qualitative interpretivist case study research method and the inductive research approach. We studied five communities inhabited by the indigenous peoples in Bauchi State, Northeast Nigeria. We collected data using the focus group-based interview and observation. Data collected during the course of the study was analyzed using the thematic data analysis technique. The study findings reveal that socio-economic assumptions that revolve around the indigenous peoples' assumptions about the appropriateness of interest-based loans and cybersecurity issues impacted the number of mobile financial services situated in the research contexts, and in effect the extent of financial inclusion among the indigenous peoples. We conclude that government and organizations pushing financial inclusion among the indigenous peoples should establish measures that can be used to manage socio-cultural, economic and technological factors that that negatively impact the financial inclusion of the indigenous peoples.

**How to cite this paper:**
Abduljalal Hassan, Samuel C. Avemaria Utulu. 2022. Socio-economic and Technological Factors Influencing Financial Inclusion among Indigenous Peoples in Bauchi State, Nigeria. A Paper Presented at the *International Conference on Information and Communication Technologies and Development*, June 29 to July1, at the University of Washington in Seattle, USA.

---

* Place the footnote text for the author (if applicable) here.

# 1 INTRODUCTION

Scholars have argued that financial inclusion is pivotal to peoples' and societies' socio-cultural and economic development [1]. Financial inclusion denotes a condition in which everybody irrespective of their status, economic standing and beliefs have access to, and could use the financial products and services provided by commercial banks and other financial service institutions. Financial inclusion is taken as important because issues relating to creating conditions that enable people to function and achieve their socio-cultural and economic goals [2] and being able to manage how vulnerabilities impact their capital assets [3], all depend largely on financial inclusion. It follows that everything people need to live well, particularly those living below the poverty line, requires having access to, and spending money. The requirement makes it imperative for stakeholders to begin to reflect and work on how the financial inclusion of indigenous peoples can be assured. Given that indigenous peoples are usually secluded, basic information available about them in the extant literature are those produced by the World Bank. [4] for instance, posit that the indigenous peoples are culturally separate tribes and live based on civilizations that have links to their ancestors. They live in rural settlements and depend largely on natural resources. Maranga [5] argues that membership to any indigenous tribe is usually certified through self-identification and validation done by the indigenous peoples themselves.

Given that financial inclusion is information and communication technology (ICT) based, it usually seems that it is impossible to achieve the financial inclusion of the indigenous peoples. The ways of life of the indigenous people and the conditions under which they live makes it difficult for them to benefit from ICT driven development programs propagated by intergovernmental organizations, governments and the private sector [6] For intergovernmental organizations such as the World Bank, United Nations, African Development Bank, among others, acquiring basic information about the exact needs of the indigenous people and how to align modern development projects with their cultures, usually constitute challenges to implementing socio-cultural and economic development programs among indigenous peoples[7, 8]. For governments, particularly those in developing contexts, most socio-cultural and economic development programs they initiate adopt the top-down approach [9]. Here, a top-down development approach means initiating development programs from urban centers with the aim of expanding them to rural areas, including those inhabited by the indigenous peoples. The primary challenge private business organization face which limits their interests in setting up businesses, including commercial banks, microfinance banks, and other financial services, in settlements where the indigenous peoples live is the viability of making profits from the businesses [2, 4, 10]. Irrespective of these challenges however, efforts are being made by intergovernmental organizations, governments and the private sector to initiate financial inclusion programs in settlements inhabited by the indigenous peoples.

This study was carried out with the aim of assessing the challenges to the financial inclusion of the indigenous peoples living in Bauchi State, Nigeria. The study was necessitated by the growing popularity of the policy initiatives being implemented by both national and state governments in Nigeria to promote the financial inclusion of people in rural areas, including the indigenous peoples. The preliminary observation done in the cause of the study showed that stakeholders' expectation about the expansion of financial inclusion among the indigenous peoples living across Nigeria has not been met. To understand the factors behind this, we decided to carry out an empirical case study in Bauchi State, Nigeria. Consequently, the study was driven by the research study question: what are the challenges of financial inclusion among the indigenous peoples in Bauchi State, Nigeria? Given the novelty of the issues impacting on financial inclusion among indigenous people, particularly as it concerns the research study context in Bauchi State, Nigeria, the study adopted the inductive research approach. The unique thing about the inductive research approach is that it requires that research is carried out without reference to existing formal theories [11]. A more comprehensive treatment of the research approach adopted in the study was presented in the methodological assumption segment below.

# 2 LITERATURE REVIEW

The rate of growth of the body of literature on financial inclusion and the indigenous peoples is very slow. The rate is persistently slow despite that financial inclusion has become a critical public policy objective in the aftermath of the financial crisis. Financial inclusion refers to a financial system in which all economic actors, particularly those in disadvantaged regions and with low incomes, have adequate access to financial services. According to Liu, Huang [1] financial inclusion denotes having access to suitable, low-cost, fair, and secure financial products and services provided by commercial banks and other financial service providers. Kim, Yu [12] posits that financial inclusion results in an adequately functioning financial system that allows people and businesses to gain from savings and consumption, financing, risk management, and credit generation. Given the importance of financial inclusion to achieving socio-cultural and economic development goals, the Central Bank of Nigeria (CBN) introduced a financial inclusion initiative in Nigeria in 2013. Taiwo [13] reveal that the financial inclusion initiative put in place in Nigeria resulted in the increase in the number of banked citizens. They however note that the increase in the number of banked citizens has very little impact on the socio-cultural and economic development of citizens, particularly those living in rural areas. The claim was corroborated by Nwankwo and Nwankwo [6] claim that people living in rural areas across Nigeria are among the groups of people that are most financially excluded. Nwankwo and Nwankwo [6] also reveal that Nigeria ranks very poorly among other developing countries in the league of countries with required level of financial inclusion. Disappointingly, there have not been efforts made by scholars in Nigeria to study financial inclusion among the indigenous peoples in the country. Policy papers and statements usually present those living in rural areas as if they are the indigenous peoples.

Indigenous people are culturally separate tribes who live in settlements. Indigenous peoples live based on the civilizations that continuously link them to their ancestral heritages, and depend mainly on natural products and resources for food and sustenance [4]. According to the World Bank, there are between 370 and 500 million indigenous people worldwide[4]. The United Nations reveal that there is no legally acknowledged definition of indigenous people. It follows that the notion proposed by the United Nations on legal acknowledgement was based on assumptions regarding how official definitions are used to classify people across the globe mainly for immigration purposes. Many academics have, however, discussed the notion of the indigenous peoples. According to Maranga [5], an indigenous person is a person who belongs to a defined indigenous community by self-identification as indigenous and who is acknowledged and accepted by the population as one of its members. Usually, indigenous peoples are seen as indigenes, aboriginal people and locals. This is to say that indigenous peoples are linked to a distinguishable culture including, religion, beliefs, conventions, and values that succinctly separate them from other people [14]. The indigenous peoples inhabit lands and territories that belong to their ancestors, abide by rules set and administered by tribal institutions, maintain traditional livelihood systems in their ways of farming, fishing, and craftworks [15]. Maranga [5] went on to say that the features of indigenous peoples necessitate their separation from cultural groupings during socio-economic development study.

Research studies carried out to study financial inclusion and the indigenous peoples evolved as two separate themes in the extant literature. However, in the recent past, scholars have started to study financial inclusion among indigenous peoples [16-18], although the number of studies is still very small. Studies looking at both financial inclusion and the indigenous peoples were necessitated because of the relationship between socio-cultural and economic development and financial inclusion. In other words, there is a strong assumption that financial inclusion is a primary factor that comes to bear in the socio-cultural and economic development of the indigenous peoples. Socio-cultural and economic development has been defined as the ability to manage vulnerability factors such as social, human, natural, physical and financial factors and provide enabling contexts (both physical and technological) for people to function and achieve their life goal [2, 19]. A first look at financial inclusion and the indigenous peoples will give an impression that the two concepts are a good example of existential paradox. In other words, while financial inclusion is a modern approach that uses ICT to promote access to financial services and products provided by commercial banks, microfinance banks and other financial institutions, indigenous people strictly maintain their connection with traditional technologies including assumptions and techniques for making, keeping and accessing money [4, 5, 20]. The paradoxical relationship between financial inclusion and the indigenous peoples results in deep rooted challenges that make it difficult for stakeholders to implement financial inclusion initiatives among indigenous peoples across the globe.

There is an obvious dearth of studies that assessed financial inclusion among the indigenous peoples in the extant literature. Stakeholders constantly rely on studies that were reported in the extant literature on financial inclusion among people that are not indigenous peoples when trying to deal with issues related to financial inclusion among the indigenous peoples. Given that the literature is rich with studies that look into financial inclusion policy efforts made at country, regional and global levels, these studies provide the basis for reaching conclusions on how to promote financial inclusion among the indigenous peoples. Arun and Kamath [17] for instance, used macro data to look at the policy framework developed by India, South Africa, and Australia to address financial inclusion. Asuming, Osei-Agyei and Mohammed [21] studied the state and level of financial inclusion in thirty-one sub-Saharan African countries. The study reveals the influence of age, gender, education and wealth has on the level of financial inclusion among individuals. It also reveals the impact the growth rate of GDP and existence of financial institutions have on the level of financial inclusion at macroeconomic level. Another study that looked at financial inclusion in sub-Saharan African countries is Ulwodi and Muriu [22] study. The study reveals that low-income level, literacy level, age, and proximity of financial institutions are critical factors of financial inclusion in rural areas. Given the dearth of studies that look at the factors that come to bear in financial inclusion in sub-Saharan Africa, countries in the region have very little scientifically derived knowledge that could be used to make inference to situations that impact financial inclusion among indigenous peoples.

However, there are a couple of studies in the extant literature that were carried out to understand the factors that come to bear in the implementation of financial inclusion among them. Godinho and Singh [18] is a good example of this kind of study. Godinho and Singh [18] reveals that the success of financial inclusion initiatives in indigenous Australia is influenced by remoteness, cultural preferences for face-to-face banking, and the unpopularity of mobile banking in indigenous communities. In another study carried out to assess financial inclusion among the indigenous peoples, Johnson, Krijtenburg [16] reveal that differences in norms and expectations between the indigenous peoples and those in charge of financial inclusion initiatives constitute critical challenges to attaining the financial inclusion of indigenous peoples. Usually, norms and expectations are cognition based and reflect people's views and opinions. [7] argues on the importance of aligning norms and expectations of people living in indigenous communities and that of those who promote development initiatives to attain development goals at the base of the pyramid. He proposed communication and dialogue as the main tools for aligning norms and expectations held by different groups. Despite the insights provided by these studies, there is still an obvious dearth in studies that look into the factors that influence financial inclusion of the indigenous peoples in Africa, and Nigeria most especially.

## 3 METHODOLOGICAL ASSUMPTIONS

The research design adopted in the study is the qualitative research design. There are three main types of research designs that could be adopted by researchers in the social sciences namely, quantitative, qualitative and mixed method research designs [23]. We adopted the qualitative research design, which by implication means that our methodological assumptions including, research philosophy, data collection and analysis techniques and research method, are driven by the qualitative nature of our inquiry [24]. The research approach adopted for the study is the inductive research approach. In the social sciences, the dominant research approach scholars adopt are the deductive research approach and the inductive research approach. The two research approaches are distinguished by the type of knowledge that scholars use to drive their studies. For instance, it is believed that the deductive research approach permits scholars to adopt existing knowledge which is usually in the form of formal theories, while inductive research requires that scholars should carry out research studies without formal theories [25, 26]. We adopted the inductive research approach which by implication means that we did not adopt any formal theory or theoretical perspectives to inform our study. The research philosophy that we adopted for the study is the interpretive research philosophy. Interpretivism is one the three core research philosophies adopted in the social sciences with the others being, positivism and critical realism [27, 28]. The interpretivism proposes that reality, including people, social contexts and technology are socially constructed. This is to say that they exist based on the experiences, abilities and interpretations of those that are enacting their everyday life realities based on them. The research method adopted in the study is the case study research method. The case study research method informs an in-depth study of a phenomenon and provides a basis for coming up with preliminary insights on the phenomenon [29]. The interpretive case study research method therefore combines the qualities of interpretivism with that of the case study research method. The implication is that phenomena under study are taken to be fuzzy and evolving, including the people, contexts and technologies embedded in the study contexts [28, 30].

The study sample comprises indigenous peoples in sub-urban areas of Bauchi State, Northeast Nigeria. The indigenous peoples that were studied live in the areas that have become known in contemporary time as sub-urban areas. It follows that modernization and expansion of urban areas have had a dramatic impact on the indigenous peoples that were studied. The sampling procedure used for the study is the convenience sampling technique and the purposive sample technique. The convenience sampling technique and the purposive sampling technique allow scholars to select their sample population without paying attention to the statistical rigors that determine sampling in studies that adopted the qualitative design [31, 32]. Given that we adopted the convenience sampling technique and purposive sampling technique, we selected the indigenous peoples that participated in the study based on proximity and the inexcusable level of financial inclusion that we observed among them. The indigenous peoples that were sampled do not speak the English Language or any other language but the Hausa Language. They were Hausa people and have been living in Bauchi State for centuries. They inhabited Gudum, Kangere, Inkil, Tirwun, and Dumi. The data collection technique that we used in the study is the focus group. Usually, focus group discussion group discussion comprises between five and ten research samples. The implication is that focus group discussion enables researchers to have access to different views which may be complementary or contradictory from study participants [33]. We carried out six focus group discussions with the indigenous peoples that we studied. Each of the focus group discussions lasted for at least one hour. The focus group discussions were held using the Hausa Language and was carried out by the first author given that he is a traditional Hausa Language speaker and an indigene of Bauchi State. We adopted the thematic data analysis technique. The thematic data analysis technique requires that scholars develop themes from the interviews held with study participants and group the different themes to show the relationships among them [34, 35].

With regards to meeting the ethical requirements for the study, the study participants were dealt with at both individual and group levels. Because the first author could speak the Hausa Language and is an indigene of Bauchi State, he consulted with the participants on the ethical requirements directly. He explained the objective of the study and the role each of them is expected to play to them. He made it clear to the study participants that they could withdraw from the study at any time and are not obliged to provide information that they feel is personal and detrimental to their continuous existence within the community. Before each focus group discussion session, the first author and the study participants usually talk on issues not relating to the study and he uses the opportunity to familiarize himself with the group and to observe the study contexts. The research procedure is as follows. The study began with an observation of the bad state of financial inclusion among the indigenous peoples in Bauchi State. Further observation was carried out to understand the lack of financial inclusion in the study contexts. The first author visited the five research contexts on several occasions to further observe the state of financial inclusion in the research study contexts. Field notes were taken during the visitation on issues observed. Given that the study adopted the inductive research study approach, study data was collected without reference to any formal theory and/or theoretical perspectives. The focus group discussion and observation comprised the study data collection technique. Focus group discussions held during the study were recorded using a mobile phone as the recording device, while observations were recorded in the field note. The data were analyzed using the Atlas Ti software and were grouped into themes. After coming up with themes from the study data, the theoretical elaboration was carried out to relate study findings with theoretical insight available in the extant literature and to come up with two propositions. After this procedure, the writing up of the study was done.

## 4 PRESENTATION OF RESEARCH STUDY FINDINGS

The objective of the study is to identify and explain the challenges to financial inclusion among the indigenous peoples in Bauchi State, Northeast Nigeria. The study became imperative because of the disappointing state of financial inclusion among the indigenous peoples in Nigeria. Consequently, five communities in a sub-rural area in Bauchi State were studied as cases to serve as a basis to expose and explain the challenges to the financial inclusion of the indigenous peoples in Nigeria. The research findings are presented below.

### 4.1 Interest-Based Loan Assumptions, Low Number of Mobile Financial Service Agents and Financial Inclusion of the Indigenous Peoples

Cultural issues related to the assumptions held by the study participants about the appropriateness of interest-based loans was found to be a crucial challenge to financial inclusion among the indigenous peoples studied. It was revealed that the religious background of the indigenous peoples that were studied abhorred interest-based loans. Hence, funds provided through loans by both the Federal Government of Nigeria and the Bauchi State Government which could have provided the indigenous peoples the opportunity to own bank accounts was considered unacceptable practice. Usually, government loans are paid into commercial bank accounts and could help a large number of the indigenous peoples to own bank accounts with commercial banks. Owning accounts would also have helped them to have access to financial products and services provided by commercial banks in Nigeria. The assumption that interest-based loans are not appropriate had a crucial impact on the extent to which they enjoyed financial inclusion. This is given that government policies and practices surrounding implementing financial inclusion among the indigenous peoples is largely dependent on using government loans to spur bank account ownership. It follows that the indigenous peoples' religious beliefs influenced the evolution of the assumption that interest-based loans are not appropriate.

One of the participants, tagged Participant 2, explains how assumptions on the inappropriateness of interest-based loans impacted on him. Participant 2 from the focus group discussion A revealed that: *Religion played a role here in accepting banking, especially when a loan is involved. And when interest is accrued, it is not accepted religiously, which makes some people stop using banks."* The participant further stated that *"culture [and the assumptions they promote] stopped us from accepting something new. And our religion of course."* Interest-based loan assumptions did not only directly impact on the financial inclusion of the study participants, it also impacted negatively on the number of mobile financial service agents who were willing to set up businesses in the research study contexts. In Bauchi State, available mobile financial services include those set up by individuals with point-of-sale (POS) systems. It follows that POS is only useful to people who have bank accounts either for receiving money or for transferring money. Consequently, because the study participants did not have bank accounts given their assumption about interest-based loans, individuals who could have situated their POS systems businesses in the study contexts did not do so. It follows that culture induced assumptions about the appropriateness of interest-based loans negatively impacted on the number of mobile financial service agents in the research contexts and in effect, the level of financial inclusion enjoyed by the study participants.

### 4.2 Cybersecurity Concerns, Low Number of Mobile Financial Service Agents and Financial Inclusion of the Indigenous Peoples

Ordinarily, people may be misguided to think that cybersecurity issues are only applicable to urban centers given the rate and level of ICT implementation there. However, the study reveals that cybersecurity concerns also impact on the level of financial inclusion enjoyed by the study participants. Cybersecurity issues result mainly due to cybercrime perpetrated by culprits who capitalize on the cyber ability and knowledge of those using ICT based services, including ICT based financial services. The number of frauds in the financial sector in Nigeria and Bauchi State is alarming and affects the level of trust the indigenous peoples have on ICT based financial services. We gathered through observation that most of the study participants hear about frauds perpetrated on ICT platforms and this impacts negatively on the extent to which they trust financial services that are ICT based. There are many stories related to how fraudsters duped people who use ATM cards, and other forms of ICT based techniques used for accessing bank products and services. We also gathered that some members of the communities where our study took place have been duped or lost money to ATM transactions and similar services.

Accordingly, Participant 3 in focus group discussion B revealed that: *"there are issues related to fraud. It happened to me some time ago. I gave somebody my transfer password, and later on, he withdrew money from my account. It is a breach of trust."* The issue reported by Participant three above is a mirror of the closeness that may exist among the indigenous people and one of the ways foreign cultures have infiltrated them. Usually, the indigenous peoples live together and have strong trust for one another. Occurrence such as the one reported above presents an example of how the indigenous peoples' cultures are being infiltrated. Aside from the above, Participant 4 in focus group discussion B revealed that many feared that they will be defrauded, that is why they felt reluctant when it comes to accepting the programs that could enable them to have better levels of financial inclusion. Participant 4 posits that: *"There is one of my neighbors that was duped…He only saw the debit alert. So, we do not accept the use of mobile banking because of the fear."* The indigenous peoples live with the assumption that they are easily defrauded when they use IT based financial services provided by commercial banks. They believed that many people other than the banks have access to the information they provide to the banks including the amount of money that they may deposit in their accounts. Participant 3 in focus group discussion D testifies that: *"Yes… fraudsters call our people on the phone, they would call your name clearly and ask you to send your account number and your ATM card numbers that they are going to fix errors from your account and suddenly after that, you would be receiving a debit alert."* Issues relating to cybersecurity concerns also impacted negatively on the number of mobile financial services available in the study contexts.

## 5 THEORETICAL ELABORATION OF RESEARCH STUDY FINDINGS

### 5.1 Interest-Based Loan Assumptions, Low Number of Mobile Financial Service Agents and Financial Inclusion of the Indigenous Peoples

Issues relating to skepticism of some people towards interest-based loans have been addressed in the extant literature. The Islamic religion abhors and preaches against interest-based loans, consequently, Islamic countries have made frantic efforts to develop loan systems that are not interest based [36-38]. Our study confirms the inclination to reject interest-based loans in Islamic communities. What we directly confirmed in the study is that the indigenous peoples in the contexts of our study had a negative assumption about interest-based loans, and as a result did not participate in access bank loans that would have enabled them access bank products and services.

There are studies in the extant literature on the banked and unbanked among the indigenous peoples [18]. Usually, issues relating to assumptions about interest-based loans are discussed in relation to being banked or unbanked and not financial inclusion. It follows that our study opens up an important insight on the role assumptions about interest-based loan can play in the financial inclusion of indigenous peoples. The extended effect of the assumptions about interest-based loans namely, low number of mobile financial service agents is another important finding in our study. Usually, scholars study and report low numbers of mobile financial services in rural settlements and in some cases, the indigenous peoples' settlements without showing the factors that promote it [39, 40]. It follows that as far as we were able to confirm that no scholar has linked this to interest-based loan assumptions. In our study we discover that the low number of mobile financial service agents which directly impact the financial inclusion of the indigenous peoples was as a result of assumption about interest-based loans. This is owing to the low number of indigenous peoples that had bank accounts and were willing to use the mobile financial services. Many studies in the literature that argue that low number of financial services agents impact indigenous people's financial inclusion and that mobile money plays a crucial role in the accessibility and transformation of financial services in emerging countries [6]. Our study revealed the connections among assumptions about interest-based loan, low number of financial services in indigenous peoples' settlements and the financial inclusion of the indigenous peoples. Consequently, we came up with the proposition that,

**Proposition I (PI):**

**It is likely that the negative impact assumptions about interest-based loans have on the number of mobile financial services in indigenous peoples' settlements negatively influences their financial inclusion.**

### 5.2 Cybersecurity Concerns, Low Number of Mobile Financial Service Agents and Financial Inclusion of the Indigenous Peoples

Most studies reported in the extant literature on cybersecurity issues are based on cybersecurity experiences of people in urban areas. In fact, cybersecurity or simply put, security is a major factor that influences the acceptance and use of mobile financial services [1, 41]. Issues relating to cybersecurity have not been well thought of by scholars when dealing with why the expansion of mobile financial services in indigenous peoples' settlement is slow. Usually, factors such as mobile technology penetration, power supply and acceptance are usually considered as important factors [8, 42]. Cybersecurity has taken its toll on financial inclusion. Ambore, Richardson [43] listed cybersecurity as one of the concerns affecting financial services in general and financial service providers' attempts to increase financial inclusion. Consumers' pins and identities are stolen by cyber thieves, who then withdraw funds from their accounts, causing harm to the customers, financial institutions, and the economy [44]. Chang and Coppel [45] have linked how cybersecurity affects and improves economic growth resilience through mobile banking and digital payment systems. The digital financial market is still fragmented in developing economies due to many challenges cited in the literature, including weak infrastructure, lack of technical know-how about mobile phone use, illiteracy, and cybersecurity issues regarding financial services [13]. Past research has shown concerns with cyber-security and fraud despite the significance of leveraging IT in digital financial inclusion. It showed mistrust and distrust in the system [1].

Although there are very few studies that were reported in the extant literature on cybersecurity and financial inclusion of the indigenous peoples. Scholars and experts' actions seem to suggest that there are no cybersecurity threats in settlements where the indigenous peoples live. Our study however, reveals that cybersecurity issues are also causing impediments to the financial inclusion of the indigenous peoples. Both human-induced and technology-induced cybersecurity concerns were identified in the study. The increase of financial services fraudsters is worrying and it is expanding at a fast rate to indigenous peoples' settlements. There are several research studies reported in the literature that highlighted how security concerns impact financial inclusion. Our study corroborates these studies. The increasing numbers of transactions, including the few being completed by the indigenous peoples, means that stakeholders will continue to face cybersecurity issues [46, 47]. Even though how cybersecurity issues impact on the financial inclusion of the indigenous peoples has received little attention, its importance cannot be over emphasized [47].

The interesting thing about our study is that it revealed a link between cybersecurity issues and the low number of mobile financial service agents that operate in settlements where the indigenous peoples that were studied live. Our study showed that those, including members of the indigenous peoples' communities that would have set up mobile financial services in the settlements, were skeptical due to cyber security issues. This expands existing insights in the extant literature that mobile technology penetration, power supply and acceptance are the major factors that determine the setting up of mobile financial businesses [8, 10]. Our findings also corroborate insights existing in the extant literature that threats, negative impact on socio-cultural and economic life of individuals and loss of investments have affected both individuals and institutions [44, 45]. Consequently, we came up with the proposition that,

**Proposition II (PI):**

The impact of cybersecurity issues on the low number of mobile financial services in the settlements where the indigenous peoples live is likely to negatively impact their financial inclusion.

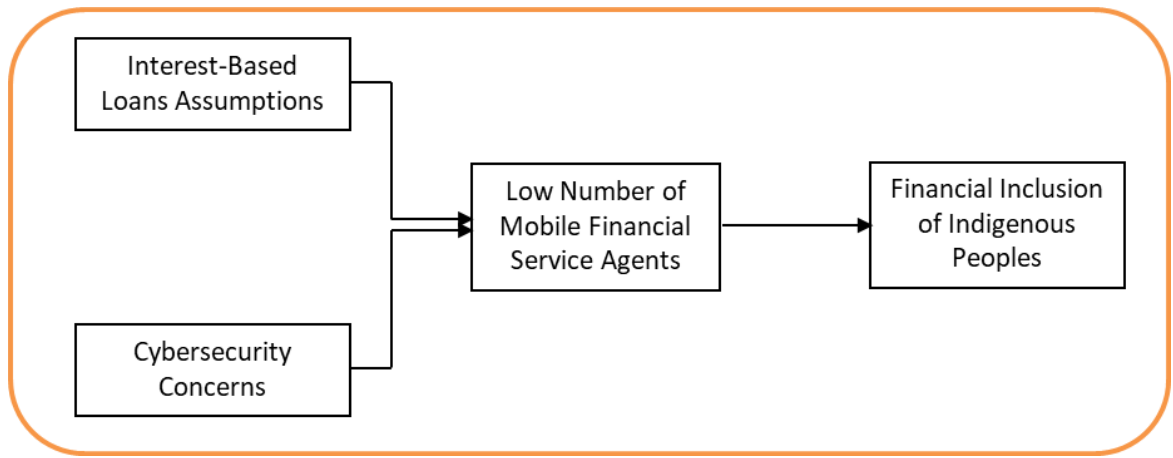

Figure 1: Factors Influencing Financial Inclusion among the Indigenous Peoples

## 6   IMPLICATIONS OF THE STUDY

The study reveals three important factors that come to play in the financial inclusion of the indigenous peoples in the study contexts in Bauchi State, Northeast Nigeria. First the study is one of the few existing studies that provide a scientifically derived knowledge about factors that come to bear in the financial inclusion of the indigenous peoples. Second, it provides important suggestions on the ways stakeholders including governments, intergovernmental organizations and the private sector can manage financial inclusion projects among indigenous peoples. It is obvious based on the study's revelation that there is a strong need to manage the assumptions indigenous peoples hold with regards to interest-based loans and cybersecurity issues and the impacts of these assumptions on the numbers of individuals and institutions that are willing to start up mobile financial service business in the settlements where the indigenous peoples live.

## 7   CONCLUSION AND LIMITATIONS OF THE RESEARCH STUDY

The study began with the objective, to find out the challenges to financial inclusion among the indigenous peoples in Nigeria. The study was necessitated by the limited scientific knowledge available in the extant literature on the factors that come to bear in financial inclusion among the indigenous peoples. This is given the importance of financial inclusion to the socio-cultural and economic development of the indigenous peoples. Our study reached its goal of eliciting the factors that come bear in the financial inclusion of the indigenous peoples. There are two primary limitations of the study namely, the extent we were able to get involved in participatory observation of the indigenous peoples that we studied. Participatory observation is useful to tease out and understand deeper cultural issues that may impact on the financial inclusion of the indigenous peoples. The second limitation is that we studied only a group of indigenous peoples that had a single cultural trait. Studies such as this, would have benefited from a cross cultural study which would have provided the basis for comparing cultures and how differences in cultures impact financial inclusion. We conclude that financial inclusion initiatives implemented among the indigenous peoples should be collaborative between those promoting and initiating the initiatives and the indigenous peoples. Our study shows that challenges to financial inclusion among indigenous peoples can be categorized into socio-cultural and technical challenges. It is important that stakeholders learn about, and pay attention to these challenges if the desire to spur socio-cultural and economic development of the indigenous peoples is to be met.

**REFERENCES**

1. Liu, G., Y. Huang, and Z. Huang, *Determinants and Mechanisms of Digital Financial Inclusion Development: Based on Urban-Rural Differences.* Agronomy, 2021. **11**(9): p. 1833.
2. Sen, A., *Inequality Reexamined*. 1992: Oxford University Press.
3. Morse, S. and N. McNamara, *Sustainable livelihood approach: A critique of theory and practice*. 2013: Springer Science & Business Media.
4. Group, W.B., *World development report 2016: Digital dividends*. 2016: World Bank Publications.
5. Maranga, K.M., *Indigenous people and the roles of culture, law and globalization: Comparing the Americas, Asia-Pacific, and Africa*. 2013: Universal-Publishers.
6. Nwankwo, O. and G. Nwankwo, *Sustainability of financial inclusion to rural dwellers in Nigeria: Problems and way forward.* Research journal of finance and accounting, 2014. **5**(5): p. 24-31.
7. Utulu, S.C.A., *Framework for Coherent Formulation and Implementation of ICT Policies for Sustainable Development at the Bottom of the Pyramid Economy in Nigeria*, in *Sustainable Development in Africa: Fostering Sustainability in one of the World's Most Promising Continents*, W. Leal Filho, R. Pretorius, and L.O. de Sousa, Editors. 2021, Springer International Publishing: Cham. p. 633-648.
8. Powell, C., *Rethinking marginality in South Africa: Mobile phones and the concept of belonging in Langa Township*. 2014: African Books Collective.
9. Semeraro, T., et al., *A Bottom-Up and Top-Down Participatory Approach to Planning and Designing Local Urban Development: Evidence from an Urban University Center.* Land, 2020. **9**(4): p. 98.
10. Timbile, A.N. and R.A. Kotey, *The Role of Financial Inclusion in Eliminating Household Poverty: Evidence from the Rural Wa West District of Ghana.* Journal of Land and Rural Studies, 2022. **10**(1): p. 75-105 DOI: 10.1177/23210249211050437.
11. Gioia, D.A., K.G. Corley, and A.L. Hamilton, *Seeking Qualitative Rigor in Inductive Research:Notes on the Gioia Methodology.* Organizational Research Methods, 2013. **16**(1): p. 15-31 DOI: 10.1177/1094428112452151.
12. Kim, D.-W., J.-S. Yu, and M.K. Hassan, *The influence of religion and social inequality on financial inclusion.* The Singapore Economic Review, 2020. **65**(01): p. 193-216 DOI: https://doi.org/10.1142/S0217590817460031.
13. Taiwo, J.N., *Effect of ICT on accounting information system and organizational performance.* European Journal of Business and Social Sciences, 2016. **5**(02) DOI: https://ssrn.com/abstract=3122462.
14. Béteille, A., *The idea of indigenous people.* Current anthropology, 1998. **39**(2): p. 187-192.
15. Kirsch, S., *Indigenous movements and the risks of counterglobalization: tracking the campaign against Papua New Guinea's Ok Tedi mine.* American ethnologist, 2007. **34**(2): p. 303-321 DOI: https://doi.org/10.1525/ae.2007.34.2.303.
16. Johnson, S., F. Krijtenburg, and K.F. Johnson S, *What do low-income people know about money?: indigenous financial concepts and practices and their implications for financial inclusion.* What do low-income people know about money?: indigenous financial concepts and practices and their implications for financial inclusion, 2014 DOI: https://hdl.handle.net/1887/54808.
17. Arun, T. and R. Kamath, *Financial inclusion: Policies and practices.* IIMB Management Review, 2015. **27**(4): p. 267-287 DOI: https://doi.org/10.1016/j.iimb.2015.09.004.
18. Godinho, V. and S. Singh. *Technology enabled financial inclusion and evidence-based policy for the underbanked: A study of remote Indigenous Australia*. in *CPRsouth8/CPRafrica2013 conference*. 2013. DOI: https://ssrn.com/abstract=2331884.
19. Serrat, O., *Knowledge solutions: Tools, methods, and approaches to drive organizational performance*. 2017: Springer Nature.
20. Olagunju, O.M. and S.C.A. Utulu, *Money market digitization consequences on financial inclusion of businesses at the base of the pyramid in Nigeria*, in *The Digital Disruption of Financial Services*. 2021, Routledge. p. 165-183.
21. Asuming, P.O., L.G. Osei-Agyei, and J.I. Mohammed, *Financial Inclusion in Sub-Saharan Africa: Recent Trends and Determinants.* Journal of African Business, 2019. **20**(1): p. 112-134 DOI: 10.1080/15228916.2018.1484209.

22. Ulwodi, D.W. and P.W. Muriu, *Barriers of financial inclusion in Sub-Saharan Africa.* Journal of Economics and Sustainable Development, 2017. **8**(14).
23. Elliott, R. and L. Timulak, *Descriptive and interpretive approaches to qualitative research.* A handbook of research methods for clinical and health psychology, 2005. **1**(7): p. 147-159.
24. Myers, M.D. and D. Avison, *Qualitative research in information systems: a reader*. 2002: Sage.
25. Siponen, M. and T. Klaavuniemi, *Why is the hypothetico-deductive (H-D) method in information systems not an H-D method?* Information and Organization, 2020. **30**(1): p. 100287 DOI: https://doi.org/10.1016/j.infoandorg.2020.100287.
26. Costal, D. and A. Olivé. *A method for reasoning about deductive conceptual models of information systems*. in *Advanced Information Systems Engineering*. 1992. Berlin, Heidelberg: Springer Berlin Heidelberg DOI: https://doi.org/10.1007/BFb0035156.
27. Burrell, G. and G. Morgan, *Sociological paradigms and organisational analysis: Elements of the sociology of corporate life*. 2017: Routledge.
28. Walsham, G., *Interpretive case studies in IS research: nature and method.* European Journal of Information Systems, 1995. **4**(2): p. 74-81 DOI: 10.1057/ejis.1995.9.
29. Ponelis, S.R., *Using interpretive qualitative case studies for exploratory research in doctoral studies: A case of information systems research in small and medium enterprises.* International Journal of Doctoral Studies, 2015. **10**: p. 535 DOI: http://ijds.org/Volume10/IJDSv10p535-550Ponelis0624.pdf.
30. Utulu, S.C.A. and O. Ngwenyama, *Rethinking theoretical assumptions of the discourses of the institutional repository innovation discipline.* 2017.
31. Suen, L.-J.W., H.-M. Huang, and H.-H. Lee, *A comparison of convenience sampling and purposive sampling.* Hu Li Za Zhi, 2014. **61**(3): p. 105.
32. Etikan, I., S.A. Musa, and R.S. Alkassim, *Comparison of convenience sampling and purposive sampling.* American journal of theoretical and applied statistics, 2016. **5**(1): p. 1-4.
33. Hennink, M.M., *International focus group research: A handbook for the health and social sciences*. 2007: Cambridge University Press.
34. Harper, D. and A.R. Thompson, *Qualitative research methods in mental health and psychotherapy: A guide for students and practitioners*. 2011: John Wiley & Sons.
35. Braun, V. and V. Clarke, *Thematic analysis*. 2012: American Psychological Association.
36. Kaleem, A. and S. Ahmed, *The Quran and Poverty Alleviation: A Theoretical Model for Charity-Based Islamic Microfinance Institutions (MFIs).* Nonprofit and Voluntary Sector Quarterly, 2009. **39**(3): p. 409-428 DOI: 10.1177/0899764009332466.
37. Bayindir, S. and M. Ustaoglu, *The issue of interest (riba) in the Abrahamic religions.* International Journal of Ethics and Systems, 2018. **34**(3): p. 282-303 DOI: 10.1108/IJOES-09-2017-0148.
38. Kabir Hassan, M. and A.Q. Aldayel, *STABILITY OF MONEY DEMAND UNDER INTEREST-FREE VERSUS INTEREST-BASED BANKING SYSTEM.* Humanomics, 1998. **14**(4): p. 166-185 DOI: 10.1108/eb018821.
39. Dass, R. and S. Pal, *Exploring the factors affecting the adoption of mobile financial services among the rural under-banked.* 2011 DOI: https://aisel.aisnet.org/ecis2011/246
40. Lema, A., *Factors influencing the adoption of mobile financial services in the unbanked population.* Inkanyiso: Journal of Humanities and Social Sciences, 2017. **9**(1): p. 37-51.
41. Hsu, C.-L., C.-F. Wang, and J.C.-C. Lin, *Investigating customer adoption behaviours in mobile financial services.* International Journal of Mobile Communications, 2011. **9**(5): p. 477-494.
42. Timbile, A.N. and R.A. Kotey, *The Role of Financial Inclusion in Eliminating Household Poverty: Evidence from the Rural Wa West District of Ghana.* Journal of Land and Rural Studies, 2021. **10**(1): p. 75-105 DOI: 10.1177/23210249211050437.
43. Ambore, S., et al., *A resilient cybersecurity framework for Mobile Financial Services (MFS).* Journal of Cyber Security Technology, 2017. **1**(3-4): p. 202-224 DOI: 10.1080/23742917.2017.1386483.
44. Hon, W.K. and C. Millard, *Banking in the cloud: Part 1 – banks' use of cloud services.* Computer Law & Security Review, 2018. **34**(1): p. 4-24 DOI: https://doi.org/10.1016/j.clsr.2017.11.005.
45. Chang, L.Y.C. and N. Coppel, *Building cyber security awareness in a developing country: Lessons from Myanmar.* Computers & Security, 2020. **97**: p. 101959 DOI: https://doi.org/10.1016/j.cose.2020.101959.

46. Badamasi, B. and S.C.A. Utulu, *Framework for Managing Cybercrime Risks in Nigerian Universities.* arXiv preprint arXiv:2108.09754, 2021 DOI: https://doi.org/10.48550/arXiv.2108.09754.
47. Khan, A., M.S. Mubarik, and N. Naghavi, *What matters for financial inclusions? Evidence from emerging economy.* International Journal of Finance & Economics, 2021 DOI: https://doi.org/10.1002/ijfe.2451.